\documentclass[a4paper,10pt,12pt]{article}
\usepackage{amsmath, amssymb, latexsym, amscd, amsthm,amsfonts,amstext}
\usepackage[mathscr]{eucal}
\usepackage{graphicx}
\usepackage{subfig}

\numberwithin{equation}{section}
\input{tcilatex}
\begin{document}

\title{Profitable forecast of prices of stock options on real market data
via the solution of an ill-posed problem for the Black-Scholes equation}
\author{Michael V. Klibanov$^{\ast }$ and Andrey V.\ Kuzhuget$^{\circ }$ \\
%EndAName
\\
$^{\ast }$Department of Mathematics \& Statistics \\
University of North Carolina at Charlotte\\
Charlotte, NC 28223.\\
$^{\circ }$ Morgan Stanley, New York, NY 10036.\\
Emails: mklibanv@uncc.edu and akuzhuget@gmail.com}
\date{}
\maketitle

\begin{abstract}
A new mathematical model for the Black-Scholes equation is proposed to
forecast option prices. This model includes new interval for the price of
the underlying stock as well as new initial and boundary conditions.
Conventional notions of maturity time and strike prices are not used. The
Black-Scholes equation is solved as a parabolic equation with the reversed
time, which is an ill-posed problem. Thus, a regularization method is used
to solve it. This idea is verified on real market data for twenty liquid
options. A trading strategy is proposed. This strategy indicates that our
method is profitable on at least those twenty options. We conjecture that
our method might lead to significant profits of those financial institutions
which trade large amounts of options. We caution, however, that detailed
further studies are necessary to verify this conjecture.
\end{abstract}

\graphicspath{
{FIGURES/}
 {pics/}}


\section{Introduction}

\label{sec:1}

The Black-Scholes equation is solved forwards in time to forecast prices of
stock options. Since this is an ill-posed problem, a regularization method
is used. Uniqueness, stability and convergence theorems for this method are
formulated. For each individual option, we use its past history as an input
data. We propose a new mathematical model for the interval of prices of the
underlying stock, as well as for initial and boundary conditions on this
interval for the Black-Scholes equation. The conventional notions of strike
prices and maturity time are not used. Our model does not make distinctions
between put and call options. Based on our results for market data, we
conjecture that our methodology might result in significant profits if
trading a large number of options, as it is done in some large hedge funds.
We caution, however, that additional studies of a substantial number of
options are necessary to verify this conjecture.

To verify the validity of our model, we use the real market data for twenty
liquid traded options. We believe that this is the best possible way for the
verification. We select options randomly among liquid ones. Those liquid
options were taken selected at http://finance.yahoo.com/options/lists/.
Market prices, implied volatility and prices of the underlying stock for our
selected options were taken from the Bloomberg terminal

http://www.bloomberg.com. The single condition we impose when selecting
these options, is that these options should be daily traded. Based on our
technique, we propose here a certain trading strategy. This strategy shows
that we are profitable on seventeen (17) out of twenty (20) options.

Our mathematical model allows one to forecast option prices of daily traded
options from the current time event to two next ones, i.e. for a short time
period. As to the time distance between two events, it depends on time units
which one is using. In our calculations one time unit is one trading day.
Thus, we forecast prices for \textquotedblleft tomorrow" and the
\textquotedblleft day after tomorrow" having knowledge of past prices of
\textquotedblleft the day before yesterday", \textquotedblleft yesterday"
and \textquotedblleft today". For our convenience, we use only last prices
in each particular day, i.e. prices which were in place just before the
closure of the market. But one unit can be one minute, one hour, one week,
etc., and also last prices can be replaced with prices at other moments of
time. An optimization of the latter is a technical issue which can be
addressed later.

In the conventional approach to the Black-Scholes equation, one solves this
equation backwards in time $t\in \left( 0,T\right) ,$ having an initial
condition at the maturity time $t=T$. The initial condition at $t=T$
includes the strike price $K,$ see Hull (2000) and Wilmott, Howison and
Dewyne (1997).\ However, there is a major drawback in this approach. Indeed, 
$T$ is usually a few months. But it is obviously impossible to forecast the
value of the volatility for such a long time period with a reasonable
accuracy. On the other hand, the solution of the Black-Scholes equation
critically depends on the volatility coefficient.

The above led us to believe that it is more natural to use the Black-Scholes
equation to forecast prices of stock options for short time periods.
However, the first major obstacle in this direction is that the problem of
solving the Black-Scholes equation forwards in time is ill-posed. Namely,
its solution usually does not exists and even if it exists, then it is
unstable with respect to the input data. Therefore, a regularization method
should be used. The second major obstacle is that some crucial input
parameters for solving that equation forwards in time are unknown. These
parameters are: the interval of prices of the underlying stock, the initial
condition, two boundary conditions and the volatility coefficient.

We positively address here the following two questions:

\textbf{Question 1}. Is it possible to forecast the option price for a
rather short time period using the Black-Scholes equation and market data?

\textbf{Question 2}. If this is possible, then can one be profitable using
this forecast and a trading strategy?

There are four main questions which need to be addressed to answer the first
question:

\begin{enumerate}
\item What is the interval for the prices of the underlying stock?

\item What are boundary and initial conditions on this interval?

\item What are the values of the volatility coefficient in the future?

\item How to solve the Black-Scholes equation forwards in time?
\end{enumerate}

We address questions 1-3 in our new mathematical model. To address the
fourth question, we use a regularization algorithm which was developed in
Klibanov (2014). We formulate theorems about stability and convergence of
this method. They were proven in Klibanov (2014) for a general parabolic
equation of the second order. The key method of proofs in these references
is the method of Carleman estimates.

We now briefly explain why the solution of the Black-Scholes equation in the
forward direction of time is an ill-posed problem. Let $\tau =const.>0$ and
the function $f\in L_{2}\left( 0,\pi \right) .$ Here is a well known example
demonstrating the ill-posedness of problems for parabolic equations with the
reversed time. Consider the following problem for the heat equation with the
reversed time, 
\begin{equation*}
u_{t}+u_{xx}=0,\left( x,t\right) \in \left( 0,\pi \right) \times \left(
0,\tau \right) ,
\end{equation*}%
\begin{equation*}
u\left( x,0\right) =f\left( x\right) ,
\end{equation*}%
\begin{equation*}
u\left( 0,t\right) =u\left( \pi ,t\right) =0.
\end{equation*}%
It is well known that this problem has no more than one solution, see, e.g.
Klibanov (2014) and Lavrentiev, Romanov and Shishatskii (1986). The unique
solution of this problem, if it exists, is%
\begin{equation*}
u\left( x,t\right) =\dsum\limits_{n=1}^{\infty }f_{n}\sin \left( nx\right)
e^{n^{2}t},
\end{equation*}%
where $\left\{ f_{n}\right\} _{n=1}^{\infty }$ are Fourier coefficients of
the function $f\left( x\right) $ with respect to $\sin \left( nx\right) .$
Consider the $L_{2}\left( 0,\pi \right) $ norm of the function $u\left(
x,t\right) $,%
\begin{equation}
\left\Vert u\left( x,t\right) \right\Vert _{L_{2}\left( 0,\pi \right)
}^{2}=\dsum\limits_{n=1}^{\infty }f_{n}^{2}e^{2n^{2}t}.  \label{1.1}
\end{equation}%
Hence, if the solution of this problem exists, then squares of Fourier
coefficients $f_{n}^{2}$ must decay exponentially with respect to $n.$
Hence, the solution of this problem exists for only a narrow set of
functions $f.$ Also, it is clear from (\ref{1.1}) that even if this series
is truncated, still small fluctuations of $f_{n}^{2}$ can lead large
variations of the function $u$. This manifests the instability of the above
problem. Furthermore, it follows from (\ref{1.1}) that the larger $\tau $
is, the more unstable this problem is. Hence, in order to obtain a rather
good accuracy, any regularization method should work only on a short time
interval. Thus, a rather accurate forecast of option prices via the
Black-Scholes equation might occur only for a rather short time period. The
latter is exactly what we are doing here.

In section 2 we present our mathematical model. In section 3 we describe our
numerical method. In section 4 we show our results for market data. We
summarize our results in section 5.

\section{The new mathematical model}

\label{sec:2}

Let $s$ be the stock price, $t$ be time, $\sigma \left( t\right) $ be the
volatility of the option. In our particular case we use only the implied
volatility listed on the market data of http://www.bloomberg.com. However,
more sophisticated models for the volatility can also be used in our model.
Let $\tau >0$ be the unit of time for which we want to forecast the option
price. In our particular case $\tau $ is one trading day since we forecast
option prices for \textquotedblleft tomorrow" having the knowledge of these
prices, as well as of other parameters for \textquotedblleft today",
"yesterday" and \textquotedblleft the day before yesterday". Since there are
255 trading days per year, then in our case $\tau =1/255.$ However, our
model can work with the case when $\tau $ is any unit of time: one hour, one
minute, etc. In any case we assume below that $\tau \in \left( 0,1/4\right)
. $ Let "today" be $t=0,$ \textquotedblleft tomorrow" be $t=\tau ,$ the day
after \textquotedblleft tomorrow" be $t=2\tau ,$ \textquotedblleft
yesterday" be $t=-\tau $ and \textquotedblleft the day before yesterday" be $%
t=-2\tau .$

We forecast only the last price of the option, i.e. the price on which one
item of that option was either bought or sold in the last trading event of a
trading day for this particular option. Let $u_{b}\left( t\right) $ and $%
u_{a}\left( t\right) $ be respectively the bid and ask prices of the option
at the moment of time $t$. Let $s_{b}\left( t\right) $ and $s_{a}\left(
t\right) $ be respectively the bid and ask prices of the stock at the moment
of time $t$. It is well known that $u_{b}\left( t\right) <u_{a}\left(
t\right) $ and that $s_{b}\left( t\right) <s_{a}\left( t\right) .$ Thus, the
market data which we need for our model are listed in Table 1. These data
are available at http://www.bloomberg.com.

%\textbf{Kirill! Vstav vertikalnue i gorizontalnye linii v tablicu 1.}

\begin{center}
\textbf{Table 1. The data we need to forecast the last option price for the
next two days }$t=\tau $ \textbf{and} $t=2\tau .$

\begin{tabular}{l|l|l|l|l|}
&  &  &  &  \\ \hline
$u_{b}\left( t\right) $ & $u_{a}\left( t\right) $ & $\sigma \left( t\right) $
& $s_{b}\left( t\right) $ & $s_{a}\left( t\right) $ \\ \hline
$t=-2\tau ,-\tau ,0$ & $t=-2\tau ,-\tau ,0$ & $t=-2\tau ,-\tau ,0$ & $t=0$ & 
$t=0$%
\end{tabular}
\end{center}

First, having discrete values of functions $u_{b}\left( t\right)
,u_{a}\left( t\right) ,\sigma \left( t\right) $ for three moments of time
listed in this table, we interpolate these functions between these three
points using the standard quadratic interpolation.\ Thus, we obtain
approximate values of these functions for $t\in \left( -2\tau ,0\right) $ as
quadratic polynomials. Next, we extrapolate functions $u_{b}\left( t\right)
,u_{a}\left( t\right) $ for $t\in \left( 0,2\tau \right) $ as those
quadratic polynomials. Since $\tau $ is small, then it is reasonable to
assume that functions $u_{b}\left( t\right) ,u_{a}\left( t\right) ,\sigma
\left( t\right) $ are approximated rather well on the time interval $t\in
\left( -2\tau ,2\tau \right) .$ Thus, we obtain functions%
\begin{equation*}
u_{b}\left( t\right) ,u_{a}\left( t\right) ,\sigma \left( t\right) \text{
for }t\in \left( 0,2\tau \right) .
\end{equation*}

Let $u=u\left( s,t\right) $ be the price of one item of the stock option.
Denote $s_{b}=s_{b}\left( 0\right) ,s_{a}=s_{a}\left( 0\right) .$ Then $%
s_{b}<s_{a}.$ Consider the numbers $u_{b}=u\left( s_{b},0\right) $ and $%
u_{a}=u\left( s_{a},0\right) .$ Let 
\begin{equation}
f\left( s\right) =\frac{u_{b}-u_{a}}{s_{b}-s_{a}}\cdot s+\frac{%
u_{a}s_{b}-u_{b}s_{a}}{s_{b}-s_{a}}  \label{2.0}
\end{equation}%
be the linear interpolation between $u_{b}$ and $u_{a}$ on the interval $%
s\in \left( s_{b},s_{a}\right) .$ Hence, $f\left( s_{b}\right) =u_{b},$ $%
f\left( s_{a}\right) =u_{a}.$

The simplest form of the Black-Scholes equation is 
\begin{equation}
Lu:=u_{t}+\frac{\sigma ^{2}\left( t\right) }{2}s^{2}u_{ss}=0,  \label{2.1}
\end{equation}%
where $L$ is the partial differential operator of the Black-Scholes
equation, see Hull (2000) and Wilmott, Howison and Dewyne (1997). We solve
this equation on the interval of stock prices $s\in \left(
s_{b},s_{a}\right) $ and for times $t\in \left( 0,2\tau \right) .$ Thus, we
impose the following initial condition at $t=0$ and boundary conditions at $%
s=s_{b},s_{a}$:%
\begin{equation}
u\left( s,0\right) =f\left( s\right) ,s\in \left( s_{b},s_{a}\right) ,
\label{2.2}
\end{equation}%
\begin{equation}
u\left( s_{b},t\right) =u_{b}\left( t\right) ,u\left( s_{a},t\right)
=u_{a}\left( t\right) ,\text{for }t\in \left( 0,2\tau \right) .  \label{2.3}
\end{equation}

Denote $Q_{2\tau }=\left\{ \left( s,t\right) :s\in \left( s_{b},s_{a}\right)
,t\in \left( 0,2\tau \right) \right\} .$ In this paper, we computationally
solve the following problem:

\textbf{Problem}. \emph{For }$\left( s,t\right) \in Q_{2\tau }$\emph{\ find
the solution }$u\left( s,t\right) $\emph{\ of equation (\ref{2.1})
satisfying the initial condition (\ref{2.2}) and boundary conditions (\ref%
{2.3}).}

This problem as well as the above interpolations and extrapolations form our
new mathematical model. Theorem 1 claims uniqueness of the solution of this
problem. This theorem follows immediately from Klibanov (2014) and
Lavrentiev, Romanov and Shishatskii (1986). Below $H^{2,1}\left( Q_{2\tau
}\right) $ and $H^{2}\left( Q_{2\tau }\right) $ are standard Sobolev spaces
of real valued functions.

\textbf{Theorem 1}. \emph{The problem (\ref{2.1})-(\ref{2.3}) has at most
one solution }$u\in H^{2,1}\left( Q_{2\tau }\right) .$

\section{Numerical method for the problem (\protect\ref{2.1})-(\protect\ref%
{2.3})}

\label{sec:3}

As it was pointed out in Introduction, the problem (\ref{2.1})-(\ref{2.3})
is ill-posed, since we are trying to solve the Black-Scholes equation
forwards rather than backwards in time,. The ill-posedness means here that
the existence of the solution is not guaranteed. In addition, the solution,
even if it exists, is unstable with respect to small fluctuations of initial
and boundary conditions, see, e.g. the book of Tikhonov, Goncharsky,
Stepanov, and Yagola (1995) for the theory of Ill-Posed Problems. Therefore,
we use the regularization method of section 5 of Klibanov (2014). In simple
terms, we find such an approximate solution of this Problem, which satisfies
conditions (\ref{2.1})-(\ref{2.3}) in the best way in the least squares
sense.

Consider the following function $F\left( s,t\right) $%
\begin{equation}
F\left( s,t\right) =\frac{u_{b}\left( t\right) -u_{a}\left( t\right) }{%
s_{b}-s_{a}}\cdot s+\frac{u_{a}\left( t\right) s_{b}-u_{b}\left( t\right)
s_{a}}{s_{b}-s_{a}}.  \label{3.1}
\end{equation}%
Then $F\in H^{2}\left( Q_{2\tau }\right) .$ It follows from (\ref{2.0}), (%
\ref{2.2}), (\ref{2.3}) and (\ref{3.1}) that%
\begin{equation}
F\left( s,0\right) =f\left( s\right) ,F\left( s_{b},t\right) =u_{b}\left(
t\right) ,F\left( s_{a},t\right) =u_{a}\left( t\right) .  \label{3.2}
\end{equation}%
Following section 5 of Klibanov (2014), consider the following Tikhonov-like
functional%
\begin{equation}
J_{\alpha }\left( u\right) =\dint\limits_{Q_{2\tau }}\left( Lu\right)
^{2}dxdt+\alpha \left\Vert u-F\right\Vert _{H^{2}\left( Q_{2\tau }\right)
}^{2},  \label{3.3}
\end{equation}%
where $\alpha \in \left( 0,1\right) $ is the regularization parameter. Note
that in the conventional case of linear ill-posed problems Tikhonov
functional is generated by a bounded linear operator, see, e.g. Ivanov,
Vasin and Tanana (2002). However, in our case $L$ is an unbounded
differential operator $L:H^{2,1}\left( Q_{2\tau }\right) \rightarrow
L_{2}\left( Q_{2\tau }\right) ,$ in which case $H^{2,1}\left( Q_{2\tau
}\right) $ is considered as a dense linear set in the space $L_{2}\left(
Q_{2\tau }\right) .$ We consider the following minimization problem:

\textbf{Minimization Problem}. \emph{Minimize the functional }$J_{\alpha
}\left( u\right) $\emph{\ in (\ref{3.3}), subject to the initial and
boundary conditions (\ref{2.2}), (\ref{2.3}).}

To solve this minimization problem computationally, we have written partial
derivatives in (\ref{3.3}) via finite differences. In particular, we have
obtained a finite difference grid covering the rectangle $Q_{2\tau }.$ Next,
we have minimized $J_{\alpha }\left( u\right) $ with respect to the values
of the function $u\left( s,t\right) $ at grid points, using the conjugate
gradient method. The starting point of this method was $u\equiv 0.$

The existence and uniqueness of the minimizer of functional (\ref{3.3})
supplied by boundary conditions (\ref{2.2}), (\ref{2.3}) is ensured by
Theorem 2. This theorem follows immediately from Theorem 5.3 of Klibanov
(2014), where the general parabolic equation of the second order was
considered.

\textbf{Theorem 2}. \emph{In (\ref{3.3}), let }$F$\emph{\ be the function
defined in (\ref{3.2}). Then for any }$\alpha \in \left( 0,1\right) $\emph{\
there exists a unique minimizer }$u_{\alpha }\in H^{2}\left( Q_{2\tau
}\right) $\emph{\ of the functional }$J_{\alpha }\left( u\right) $\emph{\
satisfying conditions (\ref{2.2}), (\ref{2.3}). Furthermore, there exists a
constant }$C=C\left( Q_{2\tau },\sigma \right) >0$\emph{\ depending only on
listed parameters such that }%
\begin{equation*}
\left\Vert u_{\alpha }\right\Vert _{H^{2}\left( Q_{2\tau }\right) }\leq 
\frac{C}{\sqrt{\alpha }}\left\Vert F\right\Vert _{H^{2}\left( Q_{2\tau
}\right) }.
\end{equation*}

To formulate the convergence result for minimizers $u_{\alpha },$ we need to
assume that there exists the \textquotedblleft ideal" exact solution of our
ill-posed problem, i.e. solution with the ideal noiseless data. Such an
assumption is one of most important building blocks of the Tikhonov
regularization theory, see, e.g. Tikhonov, Goncharsky, Stepanov, and Yagola
(1995). Thus, we assume that there exists the exact solution $u^{\ast
}\left( s,t\right) \in H^{2}\left( Q_{2\tau }\right) $ of equation (\ref{2.1}%
) with exact initial condition $f^{\ast }\left( s\right) \in H^{2}\left(
s_{b},s_{a}\right) $ in (\ref{2.2}) and exact boundary conditions $%
u_{b}^{\ast }\left( t\right) ,u_{a}^{\ast }\left( t\right) \in H^{2}\left(
0,2\tau \right) .$ We also assume that there exists a function $F^{\ast }\in
H^{2}\left( Q_{2\tau }\right) $ such that%
\begin{equation}
F^{\ast }\left( s,0\right) =f^{\ast }\left( s\right) ,F^{\ast }\left(
s_{b},t\right) =u_{b}^{\ast }\left( t\right) ,F^{\ast }\left( s_{a},t\right)
=u_{a}^{\ast }\left( t\right) ,  \label{3.4}
\end{equation}%
which is similar with (\ref{3.2}). Let $\delta \in \left( 0,1\right) $ be a
sufficiently small number. We assume that 
\begin{equation}
\left\Vert F-F^{\ast }\right\Vert _{H^{2}\left( Q_{2\tau }\right) }\leq
\delta .  \label{3.5}
\end{equation}%
Hence, it follows from (\ref{3.2}), (\ref{3.4}) and (\ref{3.5}) that the
number $\delta $ characterizes the level of the error in our data $f\left(
s\right) ,u_{b}\left( t\right) ,u_{a}\left( t\right) $ as compared with the
exact data $f^{\ast }\left( s\right) ,u_{b}^{\ast }\left( t\right)
,u_{a}^{\ast }\left( t\right) .$

In Theorem 3 we estimate the convergence rate of our minimizers $u_{\alpha }$
to the exact solution $u^{\ast },$ assuming that $\delta \rightarrow 0.$
Theorem 3 follows immediately from Theorem 5.4 of Klibanov (2014), which was
proven for a general parabolic operator of the second order. As it is always
done in the regularization theory, we choose in Theorem 3 the regularization
parameter $\alpha =\alpha \left( \delta \right) $ depending on the level of
the error in the data. Note that assumption (\ref{3.5}) of the small
perturbation in the data seems to be close to the reality.\ Indeed, although
we do not assume that the function $f^{\ast }\left( s\right) $ is linear and
functions $u_{b}^{\ast }\left( t\right) ,u_{a}^{\ast }\left( t\right) $ are
quadratic ones, as our functions $f\left( s\right) ,u_{b}\left( t\right)
,u_{a}\left( t\right) $ are, still since intervals $\left(
s_{b},s_{a}\right) $ and $\left( 0,2\tau \right) $ are small, then it is
reasonable to assume, as we do, that the function $f\left( s\right) $ is a
linear and functions $u_{b}\left( t\right) ,u_{a}\left( t\right) $ are
quadratic.

\textbf{Theorem 3}. \emph{Assume that conditions (\ref{3.4}) and (\ref{3.5})
hold. Let }$a\in \left( 0,1\right) $\emph{\ be a number, the function }$%
\sigma \left( t\right) \in C^{1}\left[ 0,a\right] $\emph{\ and there exist
constants }$\sigma _{0},\sigma _{1}>0,\sigma _{0}<\sigma _{1}$\emph{\ such
that }$\sigma \left( t\right) \in \left[ \sigma _{0},\sigma _{1}\right] $%
\emph{\ for }$t\in \left[ 0,a\right] .$\emph{\ Choose the regularization
parameter }$\alpha =\alpha \left( \delta \right) =\delta ^{2\beta },$\emph{\
where }$\beta =const.\in \left( 0,1\right) .$\emph{\ Then there exists a
sufficiently small number }$\tau _{0}=\tau _{0}\left( \sigma _{0},\sigma
_{1},\left\Vert \sigma \right\Vert _{C^{1}\left[ 0,a\right] }\right) \in
\left( 0,a\right] $\emph{\ depending only on listed parameters such that if }%
$\tau \in \left( 0,\tau _{0}\right) ,$\emph{\ then the following convergence
rates hold}%
\begin{equation}
\left\Vert \partial _{s}u_{\alpha \left( \delta \right) }-\partial
_{s}u^{\ast }\right\Vert _{L_{2}\left( Q_{\tau }\right) }+\left\Vert
u_{\alpha \left( \delta \right) }-u^{\ast }\right\Vert _{L_{2}\left( Q_{\tau
}\right) }\leq B\left( 1+\left\Vert u^{\ast }\right\Vert _{H^{2}\left(
Q_{2\tau }\right) }\right) \delta ^{\gamma },  \label{3.6}
\end{equation}%
\begin{equation}
\left\Vert u_{\alpha \left( \delta \right) }\left( s,2\tau \right) -u^{\ast
}\left( s,2\tau \right) \right\Vert _{L_{2}\left( s_{b},s_{a}\right) }\leq 
\frac{B}{\sqrt{\ln \left( \delta ^{-1}\right) }}\left( 1+\left\Vert u^{\ast
}\right\Vert _{H^{2}\left( Q_{2\tau }\right) }\right) ,  \label{3.7}
\end{equation}%
\emph{where the number }$\gamma =\left( \beta \ln 2\right) /\ln 4$\emph{\
and the constant }$B=B\left( a,Q_{2\tau },\sigma \right) >0$\emph{\ depends
only on listed parameters.}

It is clear from this theorem that the accuracy estimate (\ref{3.6}) in a
smaller domain $Q_{\tau }\subset Q_{2\tau }$ is of the H\"{o}lder type. On
the other hand the estimate (\ref{3.7}) on the upper side of the rectangle $%
Q_{2\tau }$ is of the logarithmic type. Clearly, the accuracy guaranteed by (%
\ref{3.6}) is better than the accuracy guaranteed by (\ref{3.7}). Hence, one
can expect to obtain more accurate computational results in $Q_{\tau }$, as
compared with those in $Q_{2\tau }.$ This is exactly the reason why we have
chosen to solve the problem (\ref{2.1})-(\ref{2.3}) in the larger domain $%
Q_{2\tau }.$ It is also clear from Theorem 3 why we have chosen the number $%
\tau >0$ to be sufficiently small.

\section{Results for the market data}

\label{sec:4}

In our computations we have always used $\tau =1/255,$ as mentioned in
section 2. In other words, we forecasted option prices for two days ahead,
i.e. for $t\in \left( 0,2\tau \right) $. To make sure that we obtain
accurate results at least for the computationally simulated data, we have
solved first the above Minimization Problem for synthetic data.\ To simulate
these data computationally, we have solved equation (\ref{2.1}) downwards in
time with boundary conditions (\ref{2.3}) and the initial condition at $%
t=2\tau .$ This way we have obtained the function $u\left( s,0\right)
=f_{sim}\left( s\right) .$ Next we have solved equation (\ref{2.1}) with
boundary conditions (\ref{2.3}) and the initial condition $f_{sim}\left(
s\right) $ upwards in time by the above described method. Next, we have
compared the resulting solution with the computationally simulated one.
Results were quite accurate ones. In this study of computationally simulated
data we have found that $\alpha =0.01$ is the optimal value of the
regularization parameter and we have used $\alpha =0.01$ in all follow up
computations.

Next, we have used the market data which we took at the Bloomberg terminal

http://www.bloomberg.com. We have compared our results with the true last
prices $u_{l}\left( \tau \right) $ and $u_{l}\left( 2\tau \right) $. Let $%
s_{m}=\left( s_{b}+s_{a}\right) /2$ be the mid point between bid and ask
\textquotedblleft today's" prices of the underlying stock. First, we have
calculated the minimizer $u_{\alpha }\left( s,t\right) $ via the above
minimization procedure, where $\left( s,t\right) \in Q_{2\tau }$.\ This was
done for every option for all days of its existence.

Those cases when $u_{\alpha }\left( s_{m},\tau \right) $ was between our
extrapolated values of bid $u_{b}\left( \tau \right) $ and ask $u_{a}\left(
\tau \right) $ prices, i.e. within the bid/ask spread, were not of any
interest to us and we have not traded options in those days. However, we
were only interested in those cases when our predicted price was larger than
our extrapolated ask price for at least \$0.02, i.e. $u_{\alpha }\left(
s_{m},\tau \right) \geq u_{a}\left( \tau \right) +0.03.$ Indeed, bid and ask
prices are close to each other, which makes the case $u_{b}\left( \tau
\right) \leq u_{\alpha }\left( s_{m},\tau \right) \leq u_{a}\left( \tau
\right) $ not interesting for trading. On the other hand, the cut-off value
of \$0.02 was chosen due to our computational experience. The same was for $%
u_{\alpha }\left( s_{m},2\tau \right) ,$ see our trading strategy below. In
all cases of trading strategy listed below we buy and sell options just
before the market closure.

\begin{center}
\textbf{Trading Strategy:}
\end{center}

\begin{enumerate}
\item Suppose that $u_{\alpha }\left( s_{m},\tau \right) \geq u_{a}\left(
\tau \right) +\$0.03$ and also that $u_{\alpha }\left( s_{m},2\tau \right)
\geq u_{a}\left( 2\tau \right) +\$0.03.$ Then we buy two (2) items of that
stock option at $t=0,$ i.e. \textquotedblleft today"$.$ Next, we sell one
item at $t=\tau $ and sell the second item at $t=2\tau .$In doing so, we do
not make any forecast at $t=\tau .$ Next, we apply our forecasting procedure
at the day $t=2\tau $ as above and repeat the use of our trading strategy.

\item Suppose that $u_{\alpha }\left( s_{m},\tau \right) \geq u_{a}\left(
\tau \right) +\$0.03$ but $u_{\alpha }\left( s_{m},2\tau \right)
<u_{a}\left( 2\tau \right) +\$0.03.$ Then we buy one (1) item of that stock
option at $t=0.$ Next, we sell that item at $t=\tau .$ Next, we apply our
forecasting procedure at the day $t=\tau $ as above and repeat the use of
our trading strategy.

\item Suppose that $u_{\alpha }\left( s_{m},\tau \right) <u_{a}\left( \tau
\right) +\$0.03,$ but $u_{\alpha }\left( s_{m},2\tau \right) \geq
u_{a}\left( 2\tau \right) +\$0.03.$ Then we buy one (1) item of that stock
option at $t=0.$ Next, we sell that item at $t=2\tau .$ Next, we apply our
forecasting procedure at the day $t=2\tau $ as above and repeat the use of
our trading strategy. However, we do not apply our forecasting procedure at
the day $t=\tau .$

\item Suppose that $u_{\alpha }\left( s_{m},\tau \right) <u_{a}\left( \tau
\right) +\$0.03$ and also that $u_{\alpha }\left( s_{m},2\tau \right)
<u_{a}\left( 2\tau \right) +\$0.03.$ Then we neither buy nor sell this
option \textquotedblleft today". Note that a particular case of this is the
situation when both forecasted prices for $t=\tau $ and $t=2\tau $ are
within the bid/ask spread, i.e. when $u_{b}\left( \tau \right) \leq
u_{\alpha }\left( s_{m},\tau \right) \leq u_{a}\left( \tau \right) $ and
also $u_{b}\left( 2\tau \right) \leq u_{\alpha }\left( s_{m},2\tau \right)
\leq u_{a}\left( 2\tau \right) .$
\end{enumerate}

We have evaluated twenty (20) liquid options, which were selected randomly
among those options which are daily traded (see\ Introduction). Their short
codes and our numbers for them are given in Table 2.\textbf{\ }

\begin{center}
\textbf{Table 2. Short option codes. \textquotedblleft C" and
\textquotedblleft P" mean call and put options respectively.
\textquotedblleft Number of days" means the total number of days an option
was evaluated by our procedure. }

\begin{tabular}{r|c|r}
&  &  \\ \hline
Option Number & Option Short Code & Number of days \\ \hline
1 & INTC US 01/17/15 C25 Equity & 181 \\ \hline
2 & WFC US 11/22/14 C50 Equity & 52 \\ \hline
3 & PFE US 11/22/14 C29 Equity & 68 \\ \hline
4 & YHOO US 01/17/15 C50 Equity & 267 \\ \hline
5 & AIG US 11/22/14 C55 Equity & 155 \\ \hline
6 & AAPL US 01/17/15 C80 Equity & 130 \\ \hline
7 & SBUX US 12/20/14 C80 Equity & 30 \\ \hline
8 & AAL US 01/17/15 C40 Equity & 195 \\ \hline
9 & QCOM US 12/20/14 C80 Equity & 31 \\ \hline
10 & MSFT US 01/17/15 C45 Equity & 267 \\ \hline
11 & QQQ US 11/22/14 C106 Equity & 45 \\ \hline
12 & MRK US 11/22/14 C60 Equity & 58 \\ \hline
13 & DDD US 01/17/15 C45 Equity & 119 \\ \hline
14 & WMB US 02/20/15 C48 Equity & 43 \\ \hline
15 & MSFT US 03/20/15 C50 Equity & 48 \\ \hline
16 & SPY US 02/20/15 P205 Equity & 141 \\ \hline
17 & IWM US 03/20/15 P110 Equity & 115 \\ \hline
18 & EEM US 02/20/15 P39 Equity & 108 \\ \hline
19 & YHOO US 03/20/15 C55 Equity & 76 \\ \hline
20 & HYG US 03/20/15 P86 Equity & 48%
\end{tabular}%
\textbf{\ }
\end{center}

Table 3 shows total profit/loss for each option resulting from our price
forecast and the follow up application of our trading strategy. Losses are
with the \textquotedblleft $-$" sign. We did not invest any real money.
Rather, for each option, we pretended that we buy and then sell the next day
only one item and only at those days which were recommended by our trading
strategy. To get results of this table, we have compared $u_{\alpha }\left(
s_{m},\tau \right) $ with\ the true last price of the next day $u_{l}\left(
\tau \right) .$ %\pagebreak

\begin{center}
\textbf{Table 3. Total gain/loss for each option of Table 2 during the
evaluation period. The last row shows the total profit for all options.}

\begin{tabular}{r|r}
&  \\ \hline
Option Number & Profit/loss \\ \hline
1 & 2.69 \\ \hline
2 & 0.51 \\ \hline
3 & 0.51 \\ \hline
4 & 2.93 \\ \hline
5 & 1.79 \\ \hline
6 & 4.72 \\ \hline
7 & 0.13 \\ \hline
8 & 2.65 \\ \hline
9 & 0.65 \\ \hline
10 & 1.74 \\ \hline
11 & 0.008 \\ \hline
12 & 0.9 \\ \hline
13 & $-$1.66 \\ \hline
14 & $-$0.63 \\ \hline
15 & 0.37 \\ \hline
16 & 0.98 \\ \hline
17 & $-$2.26 \\ \hline
18 & 2.9 \\ \hline
19 & 1.06 \\ \hline
20 & 0.77 \\ \hline
Total & 20.8%
\end{tabular}
\end{center}

The next interesting question is: \emph{How accurate are we in our forecast?}
Table 4 gives the average relative error of our forecast for each option.
For each option, we have calculated the relative error for those days when
we were selling that option by the above strategy. That relative error was 
\begin{equation}
\frac{\left\vert u_{\alpha }\left( s_{m},\tau \right) -u_{l}\left( \tau
\right) \right\vert }{u_{l}\left( \tau \right) }.  \label{4.1}
\end{equation}%
Next, we took the average value of numbers (\ref{4.1}) for each option of
Table 2. %\pagebreak

\begin{center}
\textbf{Table 4. The average relative accuracy of our forecast.\ The last
row shows the total average relative accuracy over all options.}

\begin{tabular}{r|r}
&  \\ \hline
Option Number & Relative Error \\ \hline
1 & 0.179935747 \\ \hline
2 & 0.282359905 \\ \hline
3 & 0.317038779 \\ \hline
4 & 0.197047078 \\ \hline
5 & 0.201012726 \\ \hline
6 & 0.217644086 \\ \hline
7 & 0.295314034 \\ \hline
8 & 0.243892569 \\ \hline
9 & 0.270193217 \\ \hline
10 & 0.237299767 \\ \hline
11 & 0.243110472 \\ \hline
12 & 0.172772735 \\ \hline
13 & 0.173193171 \\ \hline
14 & 0.371006086 \\ \hline
15 & 0.135316766 \\ \hline
16 & 0.214612849 \\ \hline
17 & 0.27188249 \\ \hline
18 & 0.258101445 \\ \hline
19 & 0.22011074 \\ \hline
20 & 0.245692644 \\ \hline
Average for All & 0.237376865%
\end{tabular}
\end{center}

Table 4 shows that the average relative error (\ref{4.1}) is rather large:
23.7\%. We have observed that we often overestimated option prices.
Nevertheless our trading strategy enables us to be profitable at least on
twenty options of Table 2.

\section{Summary and conclusions}

We have proposed a new mathematical model for the Black-Scholes equation.
Instead of traditionally solving this equation backwards in time, which is a
well posed problem, we solve it forwards in time, which is an ill-posed
problem. The theory, which was previously developed in Klibanov (2014),
guarantees the existence and uniqueness of regularized solution of this
problem as the convergence of regularized solutions to the correct one as
the level of the noise in the data $\delta $ and the regularization
parameter $\alpha \left( \delta \right) $ tend to zero (Theorems 2,3).

The best way to verify a mathematical model is to apply it to the real
market data. And this is what we have done here. We have chosen randomly
twenty (20) liquid options. Next, we have presented forecasts of option
prices for next two days for them. We have also developed a simple trading
strategy, which is using our forecast to buy and sell options. We pretended
that we buy and sell those options (the real money were not invested). The
main conclusion, which can be drawn, is that even though our technique
almost always overestimates option prices, still we got profits in seventeen
(17) of above options and we got losses in three (3) of them. Furthermore,
the total result is that, summing up these profits and losses, we are still
profitable on our randomly selected twenty options.

We point out that we did not take into account the transaction cost. We
conjecture that accounting for this cost would increase the above cut-off
value of \$0.02, while leaving the conclusion about the profit intact.

Even though our profits are small, this should not be discouraging. Indeed,
we pretended to buy and then sell the next day only one-two items of an
option at a time, and this was done only for twenty (20) options.\ However,
large financial institutions buy and sell a large number of options daily.
Hence, we conjecture that such large trading might indeed be profitable if
using the technique of this paper. Still, we caution again that further
studies of a much larger number of options should be performed to figure out
how correct our conjecture is. However, we do not have enough resources to
conduct such studies.

We believe that it is possible to refine our results and probably to
increase potential profits. We are naturally concerned about the way of
refining the accuracy of our predicted prices, see Table 4. One of such ways
is to obtain more accurate values of the volatility coefficient, as compared
with the implied volatility we have used here. Next, one should solve the
problem (\ref{2.1})-(\ref{2.3}) with this more accurate volatility
coefficient. Most likely, the volatility depends on both stock price $s$ and
time $t$, $\sigma =\sigma \left( s,t\right) .$ However, to compute the
volatility, one needs to solve a very difficult coefficient inverse problem
for the Black-Scholes equation.\ This problem is far more challenging than
the one we consider here. Indeed, the solution $u=u\left( s,t,\sigma \right) 
$ of equation (\ref{2.1}) depends nonlinearly on the coefficient $\sigma .$
On the other hand, the problem, which was solved above, is linear.

The main challenge in solving nonlinear inverse problems is linked with the
fact that the conventional least squares functionals for them are non
convex, thus having many local minima and ravines, see, e.g. Isakov (2014)
for this observation. Some ideas to calculate more accurate values of $%
\sigma \left( s,t\right) $ were proposed in Bouchouev and Isakov (1997),
Bouchouev and Isakov (1999), Bouchouev, Isakov and Valdivia (2002) and
Isakov (2014). However, these publications have not used our mathematical
model. We believe that to work with our model, some modifications of
currently known globally convergent numerical methods for coefficient
inverse problems might be applied, see Klibanov (1997) (two papers),
Klibanov and Th\`{a}nh (2014) and Klibanov and Kamburg (2015) for these
methods.

\begin{center}
\textbf{Acknowledgment}
\end{center}

The authors are grateful to Mr. Kirill Golubnichij for his help with some
technical issues of creating the pdf file for this paper.

\end{document}